# Generalized spectral phase-only time-domain ptychographic phase reconstruction applied in nonlinear microscopy

**George Dwapanyin[1], Dirk Spangenberg[1,2], Alexander Heidt[2], Thomas Feurer[2], Gurthwin Bosman[1*], Pieter Neethling[1], Erich Rohwer[1]**

[1]*Laser Research Institute, Department of Physics, Stellenbosch University, South Africa*

[2]*Institute of Applied Physics, University of Bern, Sidlerstrasse 5, 3012 Bern, Switzerland.*

*[*]gwb@sun.ac.za*

**Abstract**

Nonlinear microscopy has evolved over the last few decades to become a powerful tool for imaging and spectroscopic applications in biological sciences. In this study, i$^2$PIE, a novel spectral phase control technique was implemented in order to compress broad bandwidth supercontinuum light pulses generated in an All Normal Dispersion (ANDi) Photonic Crystal Fiber (PCF). The technique, based on time-domain ptychography, is here demonstrated in a nonlinear microscopy application for the first time. The first real-world application of this technique for second harmonic generation and two photon excitation fluorescence microscopies in biological samples is presented. We further show that in our implementation, i$^2$PIE leads to improved contrast and signal-to-noise ratios in the generated images, compared to conventional compression techniques used in nonlinear microscopy.

## 1. Introduction

Nonlinear (NL) microscopy has expanded tremendously over the last few decades and has advanced non-invasive deep tissue imaging with very high spatial resolution [1], [2]. It relies on the intrinsic response of a medium upon interaction with high intensity light sources. Depending on the strength of the incident electric field and the susceptibility of the material medium, different NL effects can be measured. Common NL techniques applied to microscopy include two photon excitation fluorescence (TPEF) [3], second harmonic generation (SHG) [4], third harmonic generation (THG) [5] and coherent anti-Stokes Raman scattering (CARS) [6]. In addition to the deep tissue imaging permissible by virtue of the near infrared (NIR) wavelengths used [7] and the high spatial resolution, the different techniques offer added advantages such as reduced photodamage and photobleaching [8], 3-dimensional localized imaging, better signal-to-noise ratio (SNR) and decreased out-of-focus background [9]–[11].

In NL optical microscopy, the challenge is to optimize the illumination signal such that it is below the damage threshold of the sample while maximizing the resulting number of emitted photons from the sample [12]. Further, since the non-linear response of the sample is highly intensity dependent a stable pulse source is required [13]. This is especially so when employing point-to-point raster scanning schemes. Of course, when raster scanning a further requirement is high pulse repetition rates in order to raster scan the sample in a reasonable time frame. The integration of mode locked ultrafast (fs) NIR laser sources in microscopy has significantly advanced multiphoton microscopy as they provide pulses with high peak intensities that can drive the NL responses without ionizing the sample, while keeping the average power modest [13]. These laser sources produce highly stable trains of identical pulses, which can be arbitrarily shaped with a pulse-shaping device [14]. The high repetition rate also enables faster raster scan time and thus lower dwell times required on each spot during a measurement leading to less energy deposition. The challenge of optimizing the NL signal is two-folded. Firstly, increasing the average laser power, may lead to an increased NL signal response, but this could result in sample damage [8], [15], [16]. Therefore, in order to avoid photodamage we must limit the average laser power. However, this reduces the NL signal. Fortunately, the NL signal for SHG and TPEF scales quadratically with the incident pulse intensity, which can be varied by altering the pulse duration. Given the inverse relationship between pulse intensity and pulse duration, it is therefore possible to increase the NL signal response for a fixed average laser power through decreasing the pulse duration (pulse compression).

In order to keep the average power low, but increase the intensity, one can use a pulse broadening and compression scheme in order to decrease the pulse duration. One such scheme is to use specially designed optical fibers for highly stable spectral broadening. The resulting phase stable pulse trains can then be temporally compressed with a compensating spectral phase mask to close to the Fourier transform limit resulting in greatly decreased pulse lengths [17]–[19]. Commercial oscillators are available to provide stable seed pulses at high repetition rates after which great care must be taken in the design of the spectral broadening and compression thereafter.

The spectral broadening must have a minimal effect on pulse to pulse variation and pulse compression has to be done to an extremely high standard. A great deal of work has been done investigating the supercontinuum generation properties of photonic crystal fiber (PCF) starting in the anomalous regime [20] where there are known issues with noise performance [21]. With the advent of the All Normal Dispersion (ANDi)-PCF, highly stable supercontinuum generation can be achieved since the noise sensitive effects to which anomalous dispersion fibers are prone to, are suppressed in these fibers [22]. Recently it was shown that polarization maintaining ANDi-PCF can generate highly stable, low-noise SC pulses [23] with minimal pulse to pulse fluctuations. These broadband light sources have several advantages over traditional light sources [24] and have been used as an alternate light source in many imaging applications [25]–[28].

Pulse characterization in the focus of an objective at the sample is required for NL microscopy. Multiphoton Intrapulse Interference Phase scan (MIIPS) [29]–[31], is predominately used for this application in recent years [2], [32], [33]. MIIPS is an in-situ phase-only measurement technique that can be applied in pulse compression but is limited in its effectiveness for the measurement and compensation of higher order dispersions [34], [35]. Further the standard application of MIIPS requires multiple subsequent measurements each resulting in an incremental phase correction which is time consuming [31].

Recently, Spangenberg et al proposed and developed a new ultrafast pulse characterization technique based on the application of the implicit ptychographic iterative engine (iPIE) in the time-domain ptychography [36], [37]. This new technique involves extending the time-domain ptychography to generalized spectral phase-only transfer functions and is known as $i^2$PIE since it measures the square of a signal resulting from the application of families of known transfer functions [38]. The $i^2$PIE technique is simple to implement and takes a spectrogram as input. The spectrogram consists of measured spectra recorded by sequentially applying the aforementioned known phase-only transfer functions to an unknown object signal (the input pulse to be characterized) and recording the subsequent second harmonic spectrum. The recorded spectrogram is then passed to the iterative engine to reconstruct the spectral phase of the input pulse. Additionally, it also converges fast and reliably and can accurately determine higher order phase.

In this study, the first real-world application of $i^2$PIE in microscopy is demonstrated. A broadband SC generated in an ANDi-PCF [39] was compressed using a pulse shaper and implementing the new $i^2$PIE technique. These compressed pulses were then used in TPEF and SHG imaging. In determining the effectiveness of $i^2$PIE as an in-line phase measurement and pulse compression technique for nonlinear microscopy, we compared it to pulses compressed with the standard MIIPS. We demonstrate the improved efficiency of $i^2$PIE over MIIPS in our implementation.

## 2. Experimental setup

The schematic setup for the developed laser imaging system is shown in Fig.1 below. A tunable femtosecond titanium-sapphire laser (Spectra-Physics Tsunami) with a 13 nm bandwidth centered at 800 nm with 12.5 nJ energy pulses (80 MHz rep. rate) was used to pump an experimental polarization maintaining ANDi-PCF (NL-1050-PM-NEG, NKT Photonics) to create a broadband supercontinuum (SC) with a spectral bandwidth of 120 nm. Fig 2 shows the typical SC spectrum used in this work spanning from 720 nm to 840 nm. The ANDi-PCF output was pre-compressed using chirped mirrors (48 reflections -175 $fs^2$ GDD per reflection, DCMP175, Thorlabs). This was done to remove most of the second order dispersions resulting from the SC generation in the PCF. Furthermore, the pre-compression reduces the total amount of phase to be corrected by the 4f shaper, so that the phase difference between adjacent pixels on the SLM is less, thereby increasing the smoothness of the applied phase function and hence the fidelity of the compression. The pre-compressed SC pulse was then sent into a 4f pulse shaper arranged as outlined by Weiner et al [40], [41] and consists of two diffraction gratings (Thorlabs GR13-0608, 600 grooves/mm) and two plano-cylindrical lenses of focal lengths 25 cm (Laser Components) with a 1-D liquid crystal spatial light modulator (SLM) (JenOptik SLM 640-d) positioned at the Fourier plane of the lenses for pulse shaper-assisted spectral phase measurements and pulse compression. Pulses originating directly from the laser oscillator, referred to as the fundamental pulse, were expanded by passing it through a telescope composed of two achromatic lenses (Thorlabs AC254-050-B ($f_1$ =50 mm) and AC254-200-B ($f_2$ = 200 mm)) in order to overfill the back aperture of the imaging objective. The generated SC pulse had a diameter of c.a. 10 mm and did not require beam expansion. The microscope is a custom-built inverted microscope consisting of an Olympus plan fluorite/IR 60x/0.9 air focusing objective and an Olympus plan achromat/10x/0.25 air condenser lens. The recorded beam waist at the focus of the objective was near identical for all of the compression techniques as well as the beam directly from the laser oscillator. The average beam waist at this position was 0.99 ± 0.01 μm. MIIPS and i$^2$PIE phase measurements were performed by generating SH from a 100-μm thick beta borate (BBO) crystal placed at the focal plane with the SH signal detected in transmission using a fiber coupled spectrometer (AvaSpec-3648, Avantes).

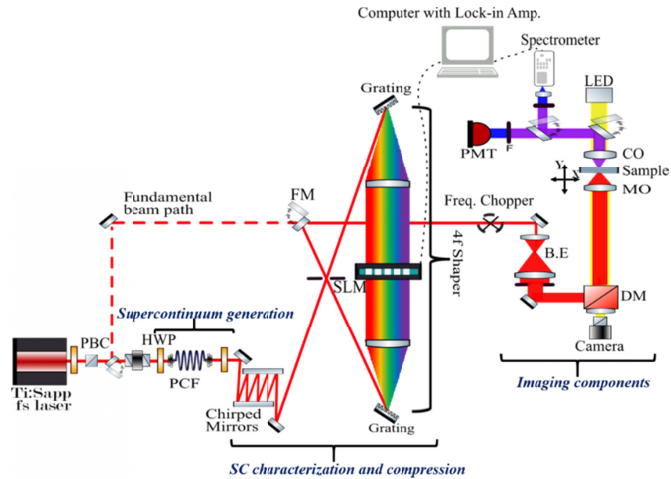

Fig 1: Integrated imaging setup showing the region of supercontinuum generation, pulse characterization and compression as well as imaging configuration. Abbreviations: PBC: polarization beamsplitter cube; HWP: halfwave plate, PCF: photonic crystal fibre; SLM: spatial light modulator; FM: flip mirror, B.E.: beam expander; DM: dichroic mirror: MO: microscope objective lens; CO: condenser lens, LED: white LED light source; PMT: photomultiplier tube.

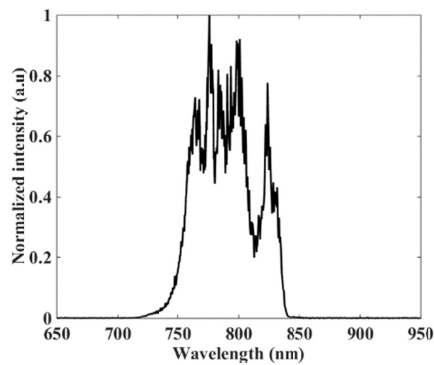

Fig 2. Typical generated broadband SC spectrum generated in an ANDi-PCF used in this work. The PCF was seeded with a pulse with central wavelength of 800 nm corresponding to an angular frequency of 2.4 rad/fs.

The custom-built imaging system performed raster scanning using computer controlled (custom Labview program) piezo stages (Newport AG-LS 25) while the signal was detected in the transmission using a photomultiplier tube (PMT) (Hamamatsu H678003) interfaced with a digital lock-in amplifier (Zurich Instrument HF2LI). Average power entering the microscope was controlled with a variable neutral density filter (NDL-25C-2, Thorlabs, fused silica). Appropriate bandpass filters were placed before the PMT to reject the transmitted laser light yet permitting the SHG and TPEF signals. A 500 Hz optical chopper generated a reference signal for the lock-in amplifier. Data acquisition was carried out using a custom-built LabView program.

### 3. Phase measurement for pulse compression

We employed the novel i$^2$PIE as a phase measurement technique for pulse compression and compared the effectiveness of this technique with the standard MIIPS compression algorithm in our implementation. Fig 3a shows the spectral phases measured with both techniques within the SC region of fig 2 (between -0.2 and 0.2 rad/fs, corresponding to 720 nm – 840 nm). It is observed that the two spectral phases have the same trend although the i$^2$PIE technique shows finer structure. A compensating negative value of the measured spectral phases was programmed onto the SLM to compress the pulse and generate near transform limited pulses. The pulse durations of the compressed pulses were recovered with a pulse shaper-assisted non-interferometric collinear autocorrelation where the carrier phase between the two replicas remains fixed and only the two envelopes are shifted and thus the measured autocorrelation traces show no oscillations[42]. Even with the time ambiguity issues of the autocorrelation function, it provided a clear indication of the pulse duration for the various compression methods. Figure 3b displays the autocorrelation function results for the SC pulse, where 'Only chirped mirror (CM) compression' represents SC compression with the GDD compensated for by the chirped mirrors. 'CM + MIIPS compression' and 'CM + i$^2$PIE compression' represents the autocorrelation of the respective techniques subsequent to CM compression. Note that the autocorrelation function displays subsidiary maxima on either sides of the main peak. These peaks are presumably due to multiple reflections in the optical system. The extracted pulse durations for the main peaks are listed in table 1.

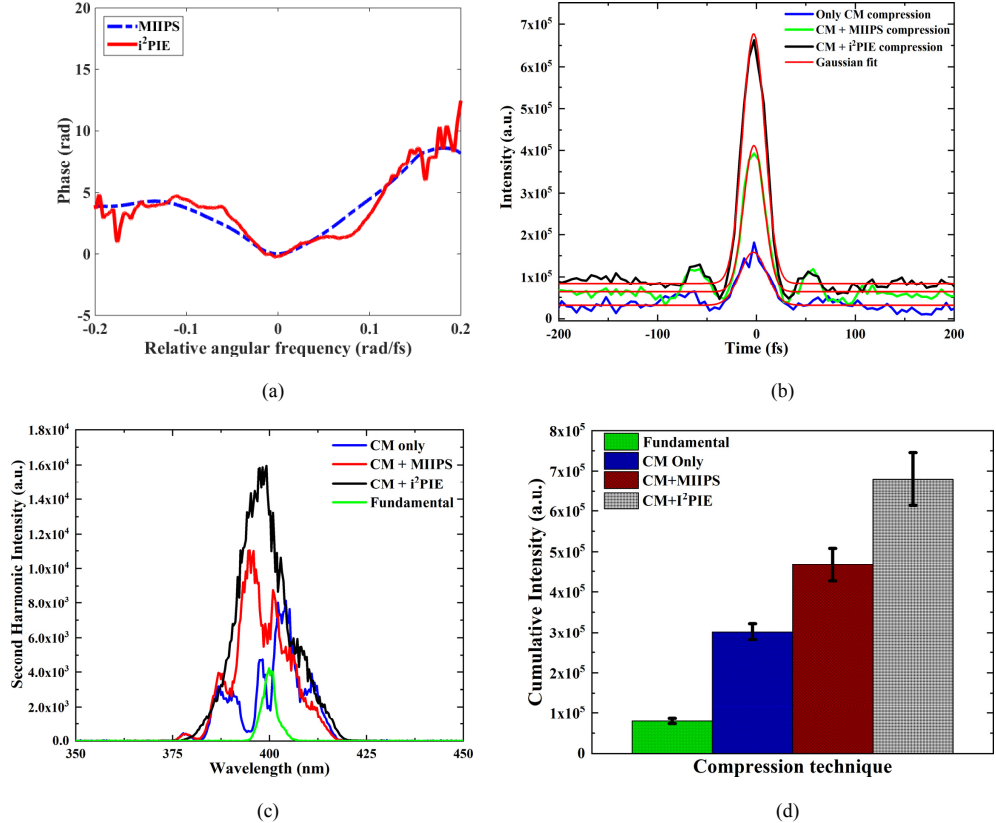

Fig 3. Spectral phases measured with MIIPS (blue) and i$^2$PIE (red) algorithms (a). Autocorrelation measurements to determine the pulse durations are plotted in (b). The SHG generated in BBO by same energy pulses for different pulse durations are shown in (c) with the integrated intensities for the entire spectral bandwidth shown in (d).

Table 1. Measured pulse durations based on compression technique.

| Technique | Duration (fs) |
|---|---|
| Chirped mirror (CM) | 25.3 ± 1.0 |
| CM + MIIPS | 21.1 ± 0.8 |
| CM + i$^2$PIE | 17.9 ± 0.7 |

Due to the setup configuration, the above-mentioned duration measurements could not be performed for the fundamental pulse (directly from the laser oscillator) as it does not traverse the SLM and hence its pulse duration at the sample was estimated to be 160 fs using the Lab2 virtual femtosecond laboratory package [43], by calculating the dispersion of the pulse through the optical setup. Fig. 3c shows the SHG spectrum generated in a BBO crystal for the fundamental pulse and the three different compressed SC-based pulses. Note that the signal recorded for the fundamental has been multiplied by a factor of 2 for clarity. The integrated intensities for the entire spectral bandwidths are plotted in Fig 3d. One can observe the expected increase in the integrated intensities for shorter pulse durations (e.g. 1.5-fold increase in SH response in the use of i$^2$PIE over MIIPS). Note that the BBO crystal used to produce these spectra did not have an appropriate thickness to adequately phase-match the entire SC spectrum. Therefore the differences in the cumulative intensities (see Fig 3d), computed via point by point addition of the spectra in Fig. 3c, provides a relative measure of the improvement in compression.

### 4. Imaging applications

The fundamental and all three SC-based generated pulses were used for nonlinear imaging of biological specimen. Second harmonic generation (SHG) imaging was carried out on the epidermal tissue of porcine skin. Fresh porcine tissue was sourced locally, ensuring that the sample was preservative free to prevent unknown contaminants from contributing to the SHG signal. No extra purification process was carried out. A thin biopsy layer was taken and placed under the microscope for SHG imaging. Since it is well known that SHG from collagen dominates the SHG response [4], [44], [45], the collagen distribution in the epidermal tissue was imaged. A 400/10 bandpass filter was used to ensure suppression of type I collagen autofluorescence which spans an emission wavelength of 300 – 500 nm [46]. The ability of the system to detect cellular level images was investigated by carrying out exogenous two photon excitation fluorescence (TPEF) imaging using fixed cultured mouse hypothalamic neuronal cells (GT1-7 cell lines). The cells were stained with Hoechst having a 350/450 nm absorption/emission maximum wavelengths and are predominately used for DNA labelling due to its lipophilic nature [47]. In working with Hoechst, the laser was tuned to a central wavelength of 780 nm (with a 13 nm bandwidth) in order to increase the two-photon absorption probability to *ca* 15% from the initial 8% at 800 nm [48]. Fig 4 shows white light micrograph images of the investigated porcine tissue and GT1-7 cell lines captured with a CMOS monochrome camera integrated in the custom-built microscope.

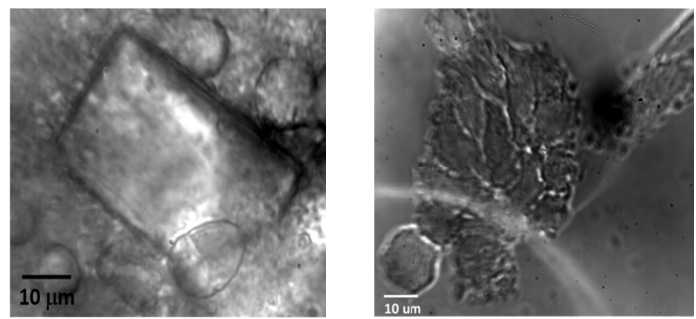

(a) Porcine epidermal tissue    (b) GT1-7 cell lines

Fig 4: Micrograph images of (a) porcine epidermal tissue, (b) fix cultured mouse hypothalamic neuronal cell lines (GT1-7 cell lines).

An average input pulse energy of 25 pJ (2 mW average input power measured just before the focusing objective) was used for all four illumination sources for both SHG and TPEF, as this power allowed for measurable signals for all four sources.

### 5. Results and discussion

Figure 5 shows the SHG images recorded in transmission for the epidermis of porcine skin while figure 6 shows the TPEF recorded in transmission for the GT1-7 cells. Please note that these images are just shown as an application as high-resolution imaging was not aim of this work. These images are presented in false colour with the relative intensities in terms of the PMT voltage response. Again, all images are plotted using the same scale. For representation purposes, the responses recorded for the scan with the fundamental pulse (directly from the laser oscillator) in fig.5a and fig 6a have been multiplied by a factor of 2 for clarity. Figures 5 (b-d) shows the SHG images recorded for only GDD compensated (b), MIIPS compressed (c) and i$^2$PIE compressed (d) pulses while figures 6 (b-d) shows TPEF images for the same order of SC pulses. There is an obvious increase in signal in the use of SC pulses over the fundamental pulse. There is also a significant improvement in the detail recorded for decreasing pulse lengths (from table 1). From figures 5 and 6 (a-d), it is observed that for the same excitation pulse energy, there is a significant increase in the SHG and TPEF responses generated using the broadband supercontinuum light sources. These imaging results highlight the advantage of SC-based sources over the use of the fundamental pulse. Furthermore, within the SC-based techniques, these results show that the enhancement in the signal responses and improvements in the amount of details recorded vary inversely with the pulse lengths recorded in table 1.

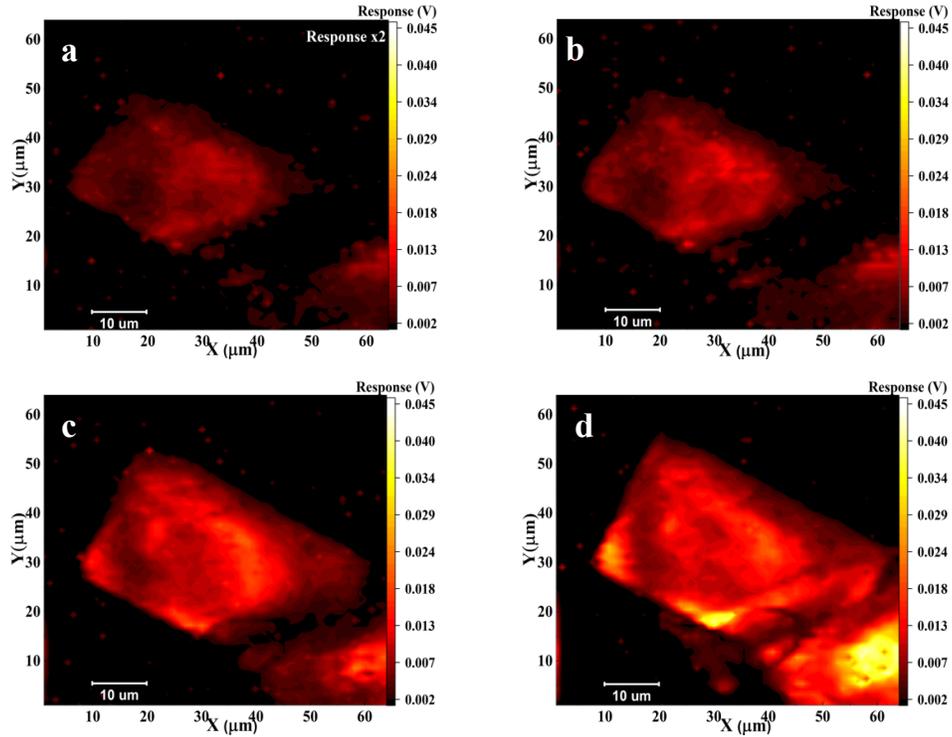

Fig 5: SHG from the epidermis of porcine skin imaged under the four configurations: (a) fundamental laser pulse (signal enhanced by a factor of 2), SC with only chirped mirror (CM) compression (b), SC with CM and MIIPS compression (c) and SC with CM and i$^2$PIE compression (d). All images taken under same conditions with pulse energy of 25 pJ. Scale bar: 10 μm.

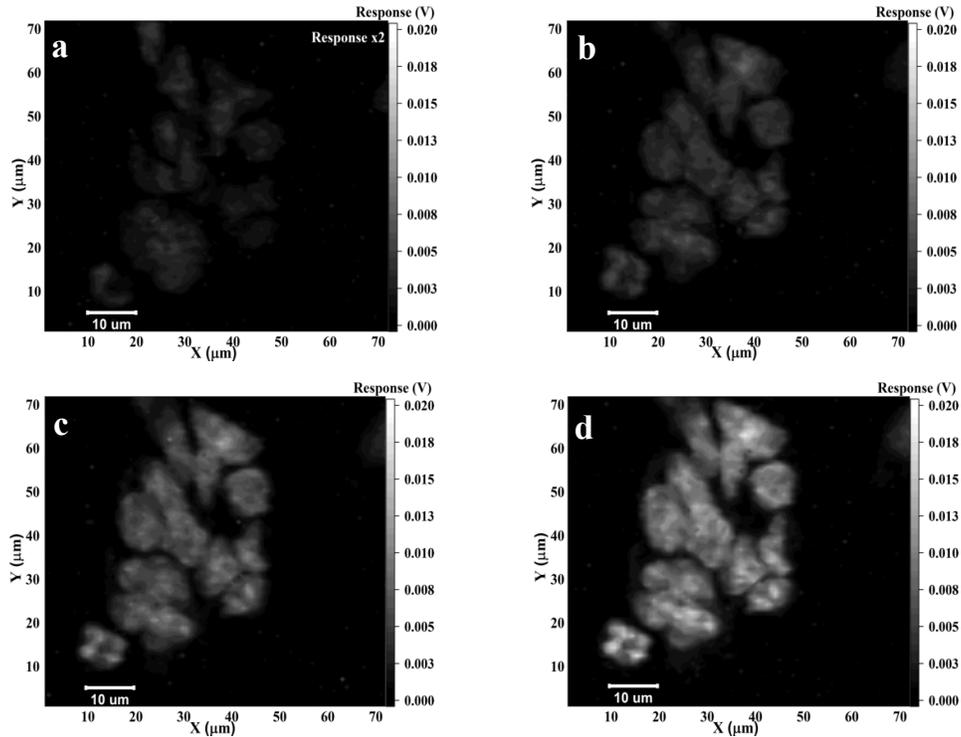

Fig 6. TPEF images from Hoechst stained GT1-7 cells imaged under the four configurations: fundamental laser (signal enhanced by a factor of 2) (a), SC with only chirped mirror (CM) compression (b), SC with CM and MIIPS compression (c) and SC with CM and i$^2$PIE compression (d). All images taken under same conditions with pulse energy of 25 pJ. Scale bar: 10 μm.

In assessing the image qualities, contrast and signal-to-noise (SNR) calculations were performed. Contrast calculations were performed using the Weber contrast [49] while SNR = maximum mean intensity / $\sigma_{noise}$ where $\sigma_{noise}$ represents the standard deviation of the noise which is the readings obtained without the sample. Note that the diffraction limited beam waist at the focus was kept to *ca* 1 μm. Given this, the calculated contrasts and SNR are presented in table 2.

Table 2. Contrast and signal-to-noise ratio measurements for SHG in dermis of porcine skin and TPEF in GT1-7 cell lines.

| Technique | Contrast | | SNR | |
|---|---|---|---|---|
| | Porcine SHG | GT1-7 TPEF | Porcine SHG | GT1-7 TPEF |
| Fundamental | 6.2± 0.7 | 8.2 ± 2.4 | 3.0 ± 0.2 | 4.2 ± 1.0 |
| Chirped mirror (CM) | 17.3± 0.9 | 35.9 ± 3.7 | 6.0 ± 0.3 | 11.5 ± 1.5 |
| CM + MIIPS | 63.3 ± 17.0 | 44.2 ± 3.2 | 16.2 ± 4.4 | 17.3 ± 1.1 |
| CM + I$^2$PIE | 135.1 ± 16.1 | 83.7 ± 9.3 | 32.3 ± 3.7 | 30.0 ± 3.2 |

The contrast recorded in fig 5d, taken with the pulses compressed with i$^2$PIE, is ≈ 2.1 times larger than that recorded for fig 5c which was taken with MIIPS compressed pulses. Similarly, there was a factor of 2.0 increase in SNR between the two images. For TPEF images taken (figure 6), the contrast (SNR) for the i$^2$PIE relative to MIIPS shows a factor of ≈ 1.9 (1.7) increase in signal responses. The signal enhancement factors for the different pulses are tabulated in table 3.

Table 3. Contrast and signal-to-noise ratio (in parenthesis) enhancement factors for Porcine SHG and GT1-7 TPEF.

| Technique | Sample | CM-only | MIIPS | i$^2$PIE |
|---|---|---|---|---|
| Fundamental | porcine SHG | 2.8 (2.0) | 10.2 (5.4) | 21.8 (10.8) |
| | GT1-7 TPEF | 4.4 (2.7) | 5.4 (4.1) | 10.2 (7.1) |
| CM-only | porcine SHG | 1.0 | 3.7 (2.7) | 7.8 (5.4) |
| | GT1-7 TPEF | | 1.2 (1.5) | 2.3 (2.6) |
| CM + MIIPS | porcine SHG | | 1.0 | 2.1 (2.0) |
| | GT1-7 TPEF | | | 1.9 (1.7) |
| CM + i$^2$PIE | | | | 1.0 |

Further comparing the images along with the calculated contrast and SNR for the MIIPS compressed pulse and the pulses with only GDD compensation i.e. only CM compressed pulses shows an increase in the MIIPS response over only GDD compensated pulses. These results are consistent with the works of Xi et al [2] and Liu et al [32] who explored these two modalities in TPEF and SHG respectively.

The higher signal strengths recorded for the i$^2$PIE over MIIPS for the same pulse energy is attributed to the i$^2$PIE providing a more accurate measure of the spectral phase of the SC pulse. This shows it is more sensitive to the complex spectral profile of the SC pulse thus making it more effective in compensating for higher order dispersions leading to better pulse compression and hence the generation of shorter temporal pulses, closer to the transform limited case. These shorter pulse durations also generate higher peak intensities and with nonlinear signal responses depending nonlinearly on the intensity of the applied field, the higher peak intensities generate higher signal responses. Due to the higher nonlinear signal responses from i$^2$PIE, lower average power (pulse energy) can be used. This leads to a decreased risk of photodamage in tissue, allowing for longer exposure/investigation time. The results obtained in SHG crystals suggest that pulses compressed with i$^2$PIE exhibit improved phase matching resulting in a better conversion efficiency across the whole spectrum. This also results in higher generated signals.

## 6. Conclusion

The first ever application of i$^2$PIE as a pulse characterization technique in SHG and TPEF microscopy has been demonstrated and shows excellent results. As a phase measurement technique that can be applied in pulse compression, i$^2$PIE has been shown to provide accurate measurement of the spectral phase of the SC pulse thus leading to a better compression compared to the standard MIIPS, in our implementation. The resultant compressed pulses provided improved contrast and signal-to-noise ratios at the same input pulse energy. Compression using i$^2$PIE results in higher peak intensities which generate stronger nonlinear signal response. This means when compared to other techniques, i$^2$PIE can easily be used with lower energy pulses whilst still providing similar signal strengths. This should lead to a reduction in photodamage in biological samples.

## 7. Funding

Portions of this research was funded by the CSIR National Laser Centre, the African Laser Centre, the National Research Foundation, "Multi-modal imaging in biophotonics: Project No. PISA-15-FSP-002 Photonics Initiative of South Africa (PISA) of the Department of Science and Technology (DST), and Schweizerischer Nationalfonds zur Förderung der Wissenschaftlichen Forschung (PCEFP2_181222, 200020-178812). Bursary assistance for George Dwapanyin was provided by the Stellenbosch Institute for Advanced Study (STIAS).

## 8. Acknowledgement

The authors would like to thank Prof Ben Loos of the Physiology department, Stellenbosch University, for providing the mouse hypothalamic neuronal cell lines (GT1-7 cells).

## 9. Disclosure

The authors declare no conflicts of interest.